\documentclass[letterpaper, 12pt]{article}

\usepackage{mathptmx}
\usepackage{url}
\usepackage[pdftex]{graphicx}
\usepackage{eso-pic}

\newenvironment{thebib}{\section*{References}\begin{list}{}{\parsep .7ex \leftmargin 1.5em \itemindent -1.5em}}{\end{list}}

\begin{document}
\setlength{\fboxsep}{10pt}
\AddToShipoutPictureFG*{\AtPageLowerLeft{
   \raisebox{2.6in}{\makebox[1.1in]{}\framebox{\parbox[t]{6.0in}{\small This article was published in {\em Statistical Applications in Genetics and Molecular Biology} on 31 October 2010 (\url{http://dx.doi.org/10.2202/1544-6115.1585}).
   Please cite as:\newline
   Phipson, B., and Smyth, G.\ K.\ (2010). Permutation $p$-values should never be zero: calculating exact $p$-values when permutations are randomly drawn. {\em Stat. Appl. Genet. Molec. Biol.} Volume  9, Issue 1, Article 39.}}
}}}

\title{Permutation $P$-values Should Never Be Zero: Calculating Exact $P$-values When Permutations Are Randomly Drawn}

\author{Belinda Phipson and Gordon K.\ Smyth\\
Walter and Eliza Hall Institute of Medical Research\\
Melbourne, Vic 3052, Australia}

\date{Published 31 October 2010; corrected 9 February 2011}

\maketitle

\begin{abstract}
Permutation tests are amongst the most commonly used statistical tools in modern genomic research, a process by which $p$-values are attached to a test statistic by randomly permuting the sample or gene labels. Yet permutation $p$-values published in the genomic literature are often computed incorrectly, understated by about $1/m$, where $m$ is the number of permutations. The same is often true in the more general situation when Monte Carlo simulation is used to assign $p$-values. Although the $p$-value understatement is usually small in absolute terms, the implications can be serious in a multiple testing context. The understatement arises from the intuitive but mistaken idea of using permutation to estimate the tail probability of the test statistic. We argue instead that permutation should be viewed as generating an exact discrete null distribution. The relevant literature, some of which is likely to have been relatively inaccessible to the genomic community, is reviewed and summarized. A computation strategy is developed for exact $p$-values when permutations are randomly drawn. The strategy is valid for any number of permutations and samples. Some simple recommendations are made for the implementation of permutation tests in practice.

\end{abstract}

\newpage

\section{Introduction}

A permutation test is a type of non-parametric randomization test in which the null distribution of a test statistic is estimated by randomly permuting the class labels of the observations.
Permutation tests are highly attractive because they make no assumptions other than that the observations are independent and identically distributed under the null hypothesis.
They are often used when standard distributional assumptions are doubtful or when the null distribution of the test statistic is simply intractable.
This article also considers Monte Carlo tests, by which we mean simulation procedures in which pseudo-random instances of a test statistic are generated from its null distribution.

The idea of permutation tests was first described by Fisher (1935), with the assumption that all possible permutations would be enumerated.
The idea of limiting the computations to a subset of permutations, randomly drawn from the set of possible permutations, was first examined by Dwass (1957).
The idea of Monte Carlo tests seems to have been first suggested by Barnard (1963).
Each of these three original papers included an expression for the exact $p$-value arising from the randomization test.
None of the expressions were suitable for practical computation at the time, however, and the results on randomly drawn permutations seem to have been largely forgotten by the literature.
Dwass (1957) in particular is not easily readable by applied statisticians, and we have not seen the exact $p$-value expression from Dwass (1957) repeated in any subsequent paper.
Even recent reviews that cite Dwass (1957) do not reproduce the exact $p$-value (Ernst, 2004).

Recent decades have seen an enormous development of bootstrap and re-sampling inference techniques (Efron, 1982; Efron and Tibshirani, 1993; Davison and Hinkley, 1997; Manly, 1997).
A key concept of this research is the idea that the distribution of an estimator or test statistic can be estimated by an empirical re-sampling distribution.
While this research has been tremendously valuable for a multitude of purposes, we believe it has had an unfortunate influence on the implementation of permutation tests.
Rather than using the exact $p$-value expressions from the original papers mentioned above, it has become common practice to use estimators of $p$-values in place of genuine probabilities.
This estimation approach is recommended in many textbooks (Higgins, 2004; Efron and Tibshirani, 1993; Good, 1994; Mielke and Berry, 2001).
Some recent papers have given confidence intervals for permutation $p$-values (Ernst, 2004), again re-inforcing the idea of $p$-value calculation as an estimation problem.
The view is often expressed that the correct $p$-values can be estimated to any required precision by drawing a sufficiently large number of permutations.

This article points out this is only true up to a point.
We show that substituting unbiased estimators for exact $p$-values often increases the type I error rate of the test.
A similar concern was raised by Onghena and May (1995) and Edgington and Onghena (2007).
We also show that, when a very large number of simultaneous tests is to be done, $p$-values can never be estimated with sufficient precision to avoid serious type I error problems at the family-wise level.
Meanwhile, exact $p$-values that do control the type I error rate correctly are readily available for little or no additional computational cost.

With the development of large genomic data sets and modern computing facilities, permutation tests have become one of the most commonly used statistical tools in modern biostatistical and genomic research.
A google scholar search for ``gene" and ``permutation test" finds nearly 8000 genomic articles, likely an underestimate of the number of relevant articles.
Permutations are almost always randomly drawn rather than exhaustively enumerated.
We were motivated to write this article because of the number of times we have seen permutation $p$-values computed incorrectly in the genomic literature.
Many published algorithms and software tools return permutation ``$p$-values" that are really estimates of $p$-values, i.e., $\hat p$-values.
In most cases, published $p$-values obtained from permutation tests are understated by about $1/m$, where $m$ is the number of permutations.
The situation is similar when Monte Carlo simulation is used to assign $p$-values.
Of particular concern is the tendency of many public software packages to return $\hat p$-values that can be exactly zero.
It is obvious that a permutation test could never yield an exactly zero $p$-value if all permutations were exhaustively enumerated, and it makes no inferential sense to assert that the $p$-value can be reduced to zero by considering only a subset of the permutations.
Most concerning of all is the potential for multiple testing procedures to be applied to $\hat p$-values that are exactly zero.
This has the potential to lead to grossly misleading results at the family-wise level, in which the results are judged to be significant at fantastically strong levels, even when all of the null hypotheses are true.

This article contains only a modest amount of novel statistical theory.
We provide a new computational solution for evaluating exact $p$-values when permutations are drawn at random.
Our main aim is to clarify the need to evaluate exact $p$-values, and to make the relevant results more accessible to the genomic community.

\section{Estimating $p$-values tends to increase the type I error rate}

Randomization tests are used in the following context.
A test statistic $t$ has been calculated from the data at hand.
We would like to use the null hypothesis tail probability $p_\infty=P(T>t)$ of the observed test statistic as the $p$-value, but this probability cannot be directly computed.
In a general sense, the purpose of randomization tests is to estimate $p_\infty$.
The purpose of this section is to point out that $p$-value calculation must take into account not just the estimate of $p_\infty$ but also the margin of error of this estimate.
Simply replacing $p_\infty$ with an unbiased estimator can result in a failure to correctly control the type I error rate.

If the test statistic $t$ is continuously distributed, then $p_\infty$ will be uniformly distributed on $(0,1)$ under the null hypothesis.
To control the type I error rate at level $\alpha$, if $p_\infty$ was known, the null hypothesis would be rejected when $p_\infty\leq \alpha$.
In other words, $P(p_\infty \leq \alpha)=\alpha$ under the null hypothesis.

Now suppose that $\hat p$ is an unbiased estimator of $p_\infty$.
It is often true that $P(\hat p\le\alpha)>\alpha$, because $\hat p<\alpha$ occurs with positive probability even when $p_\infty>\alpha$.
In general
$$P(\hat p\le\alpha)=\int_0^1 P(\hat p\le\alpha|p_\infty) dp_\infty.$$
To consider a concrete case, suppose that $m$ Monte Carlo data sets are generated under the null hypothesis, and suppose that $\hat p=B/m$ where $B$ is the number of Monte Carlo test statistics greater than or equal to $t$.
Then $m\hat p$ is binomially distributed with size $m$ and success probability $p_\infty$.
Plainly, $\hat p$ is an unbiased estimator of $p_\infty$.
Furthermore, $\hat p$ converges to $p_\infty$ as the number of Monte Carlo data sets $m\rightarrow\infty$.
This is all conditional on the observed data however.
To compute the size of the resulting Monte Carlo hypothesis test, we have to average over all possible data sets that might have been observed.
Since $m\hat p|p_\infty$ is binomial and $p_\infty$ is uniform, the unconditional distribution of $m\hat p$ is a special case of beta-binomial.
In fact, it is easy to show that $\hat p$ follows a discrete uniform distribution on the values $0,1/m,2/m,\ldots,1$:
\begin{eqnarray*}
P(\hat p=b/m) &=& \int_0^1 P(\hat p=b/m|p_\infty)d p_\infty \\
&=&\int_0^1 {m\choose b} p_\infty^b (1-p_\infty)^{m-b}dp_\infty\\
&=&\frac{1}{m+1}
\end{eqnarray*}
for all $b=0,1,\ldots,m$.
Hence the type I error rate of the Monte Carlo test that arises from using $\hat p$ as a working $p$-value is
$$P(\hat p\le\alpha)=\frac{\lfloor m\alpha\rfloor+1}{m+1}.$$
This expression can be greater or less than the nominal size $\alpha$. The discreteness of $\hat p$ however ensures that $P(\hat p \leq \alpha) > \alpha$ for most values of $\alpha$ near zero, and $P(\hat p \leq \alpha) < \alpha$ for most values of $\alpha$ near 1 (Figure A). Since $\alpha \leq 0.05$ for most experiments, the true type I error rate $P(\hat p \leq \alpha)$ exceeds $\alpha$ for most practical values of $m$ and $\alpha$. The type I error rate always exceeds $\alpha$ when $\alpha = i/m$ for some integer $i$, or when $\alpha < 1/(m+1)$. In fact $P(\hat p \leq \alpha)$ is never less than $1/(m+1)$ regardless of how stringent the desired rate $\alpha$ is chosen.

This demonstrates an unintuitive principle.
Inserting an unbiased estimator of a $p$-value for the true $p$-value does not generally lead to a test of the same size.
Rather, the type I error rate of the resulting randomized test is usually greater than that which would arise from using the true $p$-value.
To achieve a randomized test of the correct size, we must in effect substitute for the true $p$-value an estimator with positive bias.
The moral is that $p$-values must be directly evaluated as probabilities, not treated as if they were unknown parameters to be estimated in a statistical model.
\renewcommand{\thefigure}{\Alph{figure}}
\begin{figure}
\begin{center}
\includegraphics[width=3.3in,height=3.4in]{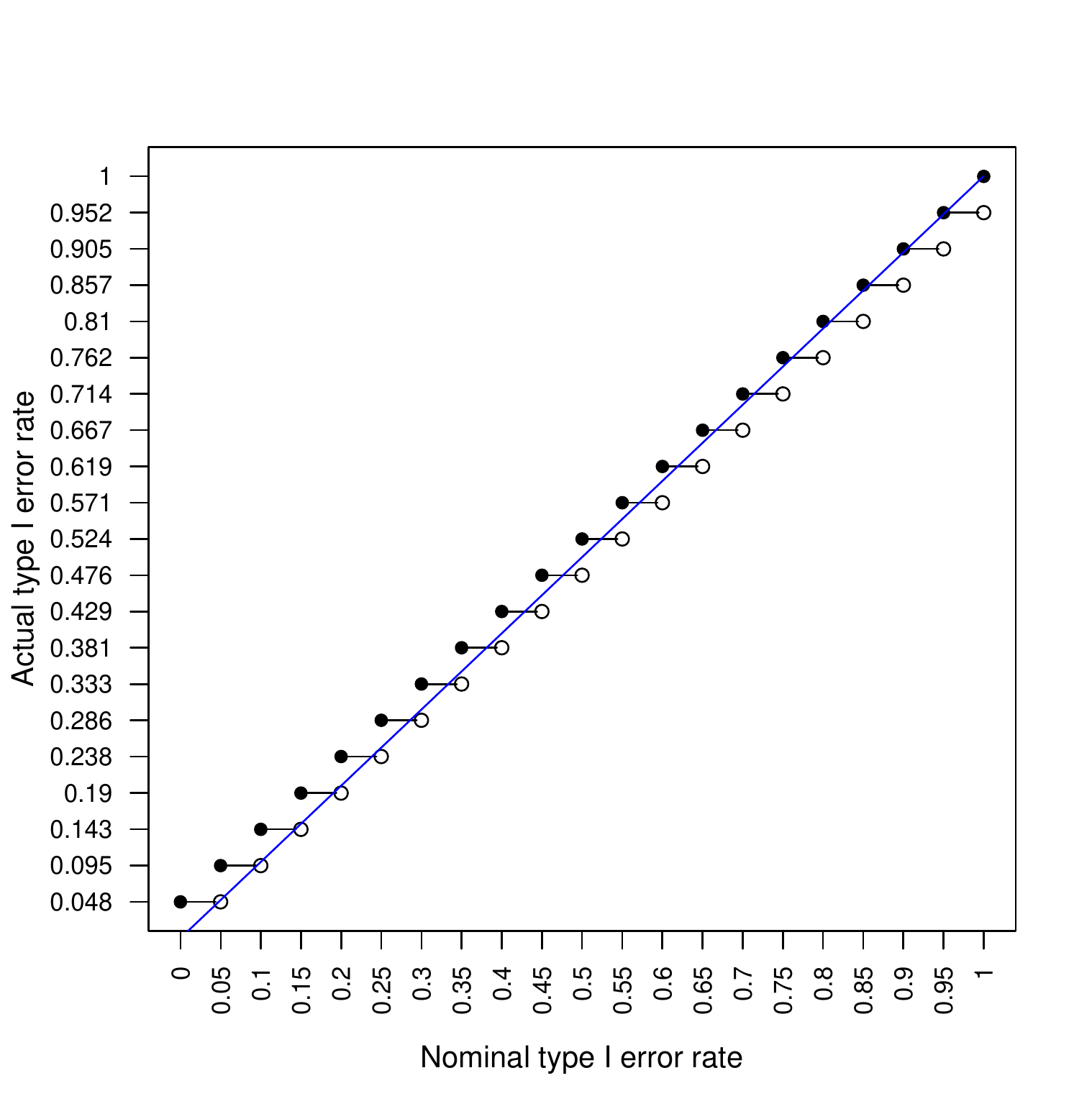}
\end{center}
\caption{Type I error rate, $P(\hat p \leq \alpha)$, arising from using $\hat p$ as a working $p$-value (y-axis) versus the nominal size $\alpha$ (x-axis) when $m=20$. The equality line, $P(\hat p \leq \alpha) = \alpha$, is shown in blue. The type I error rate is generally above the nominal rate for $\alpha<0.5$.
}
\label{fig:ns}
\end{figure}

\addtocounter{figure}{-1}

\section{Under-estimating $p$-values is dangerous in a multiple testing context}

If only one hypothesis is to be tested, mis-calculating a $p$-value probability by a small amount will often be of little consequence.
In genomic research, however, it is typically the case that many tests are to be conducted.
When the number of tests is large, any systematic under-estimation of $p$-values can lead to dangerously wrong conclusions at the family-wise level.
Suppose for example that 50,000 tests are to be conducted, one for each transcript in a transcriptome perhaps, and suppose that the resulting $p$-values are to be adjusted by Bonferroni or Holm's methods (Holm, 1979).
If each $p$-value is under-estimated by just 1/50,000, then it is highly likely that at least one null hypothesis will be rejected at any reasonable significance level, even if none of the null hypotheses are false.
In this case the family-wise type I error rate is radically wrong, even though the type I error rate is only very slightly under-estimated for each individual test.

This situation can arise easily if a permutation test is conducted for each gene on a microarray platform.
To give an example, suppose that permutation $\hat p$-values are computed from $m=1000$ sample permutations for each of 30,000 genes, where the total number of distinct permutations possible is much larger than 1000.
Suppose that a working $p$-value is computed for each gene as $\hat p=B/m$, where $B$ is the number of permutations yielding a test statistic at least as extreme as that observed from the original data.
Clearly $\hat p=B/m$ is an unbiased estimate of the tail probability for the test statistic under the null distribution induced by permutating samples.
Yet then the number of exact zero values will be around 30, even when there are no differentially expressed genes.
These 30 zero values will remain statistically significant regardless of multiple testing procedure or significance level applied.

The basic problem here is that the family-wise error rate depends on the relative accuracy on the $p$-values, rather than on additive errors.
For the smallest $p$-values, the error may be large in relative terms even when it is small in absolute terms.
When a non-zero $p$-value is mis-estimated as zero, the relative error is 100\%.

\section{Exact $p$-values for Monte Carlo tests}

It is helpful to consider first the case of Monte Carlo tests, which have much in common with permutations tests, but are mathematically simpler.
Suppose $t_{\rm obs}$ is the observed test statistic for a given hypothesis for a given experiment.
The basic assumption of Monte Carlo tests is that it is possible to generate independent random datasets under the null hypothesis $H_0$.
Typically, this is done by simulation using pseudo-random numbers.
Suppose $m$ such datasets are simulated under the null hypothesis, each yielding a distinct test statistic $t_{\rm sim}$.
In principle, the ideal $p$-value is
\[
p_\infty=P(t_{\rm sim}\geq t_{\rm obs})
\]
but this is unknown.
To compute $p_\infty$ exactly it would be necessary to generate an infinite number of simulated datasets, but only a finite number $m$ are available.

For simplicity, we assume that the test statistic $t$ is continuous, so that $p_\infty$ is uniform on the unit interval and all the $t_{\rm sim}$ are distinct. Let $B$ be the number of times out of $m$ that $t_{\rm sim} \geq t_{\rm obs}$.
The previous section showed that using the unbiased estimator $\hat p_\infty=B/m$ in place of $p_\infty$ leads to an invalid test that does not correctly control the type I error rate at the required level.
A more valid approach is to compute the tail probability directly for the Monte Carlo results.
From the point of view of the randomization test, the test statistic is $B$ rather than $t_{\rm obs}$, and the required tail probability is $P(B\le b)$. 
The previous section showed that, under the null hypothesis, the marginal distribution of $B$ over all possible data sets is discrete uniform on the integers from $0,\ldots,m$.
Hence the exact Monte Carlo $p$-value is
$$p_u= P(B\le b) = \frac{b+1}{m+1}.$$
In effect, the randomization test replaces the unknown tail probability $p_\infty$ with a biased estimator $(b+1)/(m+1)$, the amount of positive bias being just enough to allow for the uncertainty of estimation and to produce a test with the correct size.
This $p$-value calculation is presented in a number of textbooks on randomization tests including Davison and Hinkley (1997), Manly (1997) and Edgington and Onghena (2007).

\section{Sampling permutations without replacement}

Now we consider permutation tests, a type of re-sampling randomization test.
The basic setup for a permutation test is as follows.
We observe data pairs $(y_i,g_i)$, $i=1,\ldots,y_n$, where the $y_i$ are independent data observations and the $g_i$ are labels that classify the observations into two or more groups.
The aim of a permutation test is to test the association between $y$ and $g$.

A suitable test statistic $t$ is computed from the $(y_i,g_i)$.
This might be a two-sample $t$-statistic, for example, but it could in principle be any statistic designed to test for differences between the data groups.
Write $t_{\rm obs}$ for the observed value of the statistic.
The idea of a permutation test is to assign a $p$-value to $t_{\rm obs}$ by randomly permuting the class labels $g_i$ a large number of times.
Then the empirical distribution of the test statistics $t_{\rm perm}$ from the permuted data estimates the null distribution of the test statistic.
The null hypothesis $H_0$ being tested is that the $y_i$ are independently and identically distributed, versus an alternative hypothesis in which the distribution of $y_i$ differs between groups.

For simplicity, we will assume that the $y_i$ are all distinct, so that permutation yields distinct values of the test statistic unless the partition of $y$ values into groups is identical.
We assume that there are $m_t$ possible distinct values of the test statistic, not counting the original value, and that all are equally likely to arise under the permutation distribution.

The original idea of permutation tests was that all possible permutations would be enumerated (Fisher, 1935).
However $m_t$ can be very large indeed for moderate to large sample sizes, so that complete enumeration may be impractical.
In this situation, it is common practice to examine a random subset of the possible permutations.
It is important to distinguish two possible scenarios.
The random permutations may be drawn without replacement, so that all permutations are distinct.
Or permutations may be drawn with replacement, so that the same permutation may be used more than once.
We consider the first scenario now.
The second scenario is the topic of the next section.

Suppose that a random sample $m\le m_t$ permutations is drawn, without repetition, such that each permutation yields a distinct test statistic $t_{\rm perm}$, different from the original statistic.
Then the situation is essentially the same as for Monte Carlo tests considered in the previous section.
The argument used in the previous section to derive the Monte Carlo $p$-value applies again, showing that the exact permutation $p$-value is $p_u=(b+1)/(m+1)$, where $b$ is the number of $t_{\rm perm}$ greater than $t_{\rm obs}$.

This result includes the situation in which all $m_t$ possible distinct permutations are evaluated.
The exhaustive permutation $p$-value is $p_t=(b_t+1)/(m_t+1)$, where $b_t$ is the total number of distinct statistic values exceeding $t_{\rm obs}$.

\section{Sampling permutations with replacement}

\subsection{Exact $p$-values}

We now consider the case in which permutations are randomly drawn with replacement.
Each permutation is randomly generated independently of previous permutations or of the original data configuration.
This is by far the most common way that permutation tests are used in practice, especially in the genomic literature.
This scenario requires careful treatment from a mathematical point of view, because of the possibility that the random permutations may include repetitions of the same permutations and indeed of the original data.

Suppose that an independent random sample of $m$ permutations is drawn with replacement.
The resulting test statistics $t_{\rm perm}$ may include repeat values, including possibly the original observed value $t_{\rm obs}$.
The exact $p$-value is now slightly less than $(b+1)/(m+1)$, because of the possibility that the original data is included at least once as one of the random permutations.

Write $B$ for the number of permutations out of $m$ yielding test statistics at least as extreme as $t_{\rm obs}$.
We wish to evaluate the exact permutation $p$-value $p_e=P(B\le b)$ for any given $b$.
We do so using a formal conditional approach.

Write $B_t$ for the total number of possible distinct statistic values exceeding $t_{\rm obs}$, again an unknown quantity, and note $p_t=(B_t+1)/(m_t+1)$.
Although $p_t$ is in principle the ideal permutation $p$-value, it is unknown in the scenario we now consider.
If the null hypothesis is true, then $B_t$ follows a discrete uniform distribution on the integers $0,\ldots,m_t$.
Conditional on $B_t=b_t$, $B$ follows a binomial distribution with size $m$ and probability $p_t$.
Hence
\begin{equation}
p_e= \sum_{b_t=0}^{m_t} P(B\leq b | B_t=b_t) P(B_t=b_t|H_0)
=\frac{1}{m_t+1} \sum_{b_t=0}^{m_t} F(b; m,p_t)
\label{eq:permcdf}
\end{equation}
where $F$ is the cumulative probability function of the binomial distribution.
This expression is essentially the same as Proposition~1 of Dwass (1957), although Dwass did not seem to intend it for practical computational use.
Although not very difficult to derive, this exact $p$-value has not been repeated to our knowledge elsewhere in the literature.

\subsection{Computation and approximation}

Equation (\ref{eq:permcdf}) provides a formula for the exact $p$-value, but the summation may not be practical to evaluate explicitly if $m_t$ is very large.
Hence we explore an approximation that can be evaluated quickly.
Replacing the summation with an integral, and using a continuity correction, yields
\begin{eqnarray}
p_e &\approx& \int_{0.5/(m_t+1)}^1 F(b;m,p_t) dp_t \nonumber \\
&=& \int_0^1 F(b;m,p_t) dp_t - \int_0^{0.5/(m_t+1)} F(b;m,p_t) dp_t \nonumber\\
&=& \frac{b+1}{m+1} - \int_0^{0.5/(m_t+1)} F(b;m,p_t) dp_t.
\label{eq:two}
\end{eqnarray}
This shows that $p_e< p_u=(b+1)/(m+1)$, and also that $p_e$ converges to $p_u$ as $m_t$ increases.
The last integral in equation (\ref{eq:two}) is easily evaluated using numerical integration algorithms.
We have found it convenient to use Gaussian quadrature.

The valid but conservative $p$-value $p_u$ has been recommended by a number of authors including Davison and Hinkley (1997), Manly (1997) and Ernst (2004).
The extent to which $p_u$ over-estimates the exact $p$-value depends on $m_t$.
When $m_t=1000$, $p_u$ over-estimates the 0.05 $p$-value by about 2\%, but smaller $p$-values are over-estimated by greater relative amounts (Figure~\ref{fig:puvspe}b).
When $m_t$ is small, $p_u$ can be very conservative (Figure~\ref{fig:puvspe}a).

\renewcommand{\thefigure}{\arabic{figure}}

\begin{figure}[!ht]
\includegraphics[width=2.85in,height=2.85in]{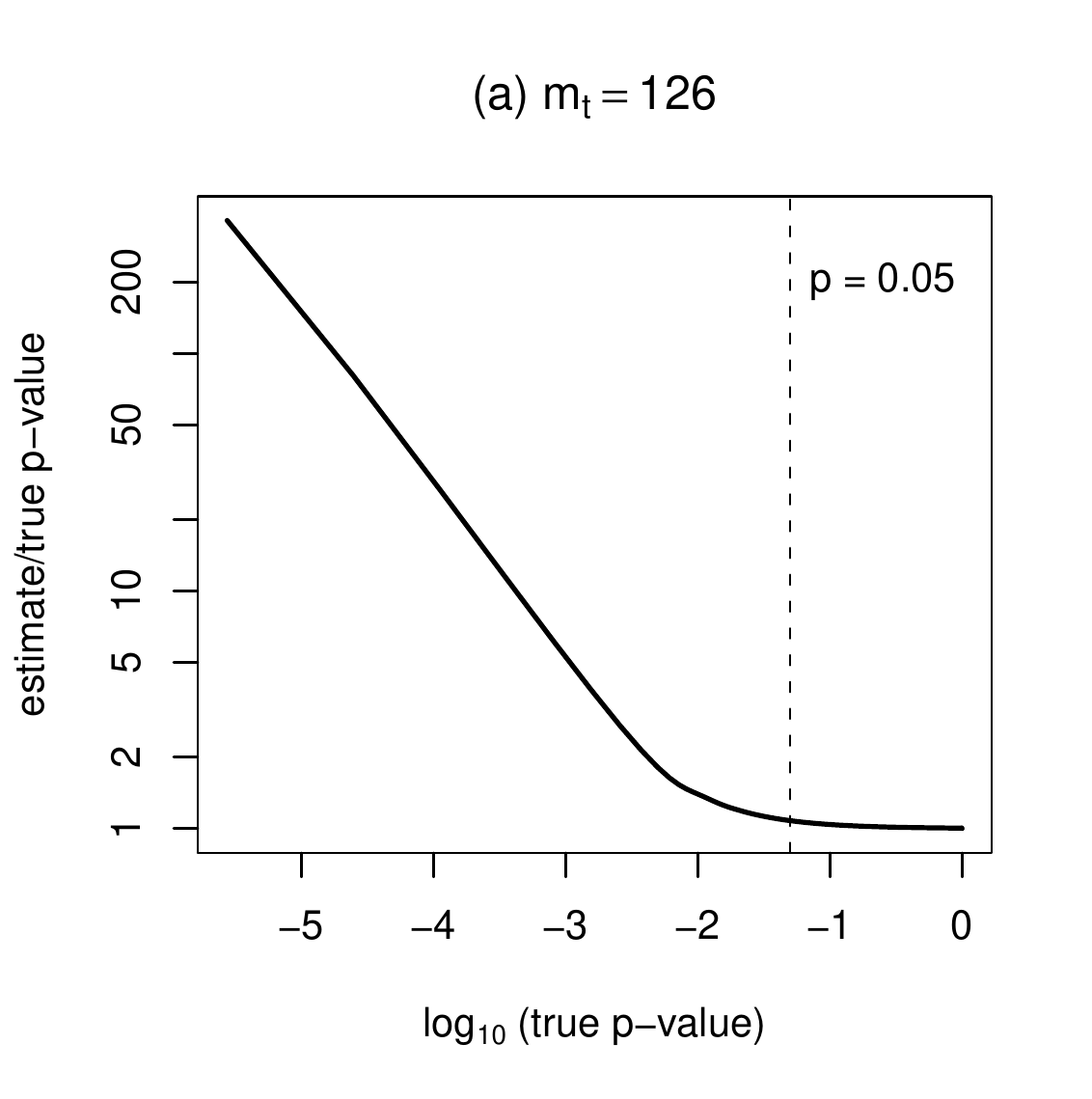}
\includegraphics[width=2.85in,height=2.85in]{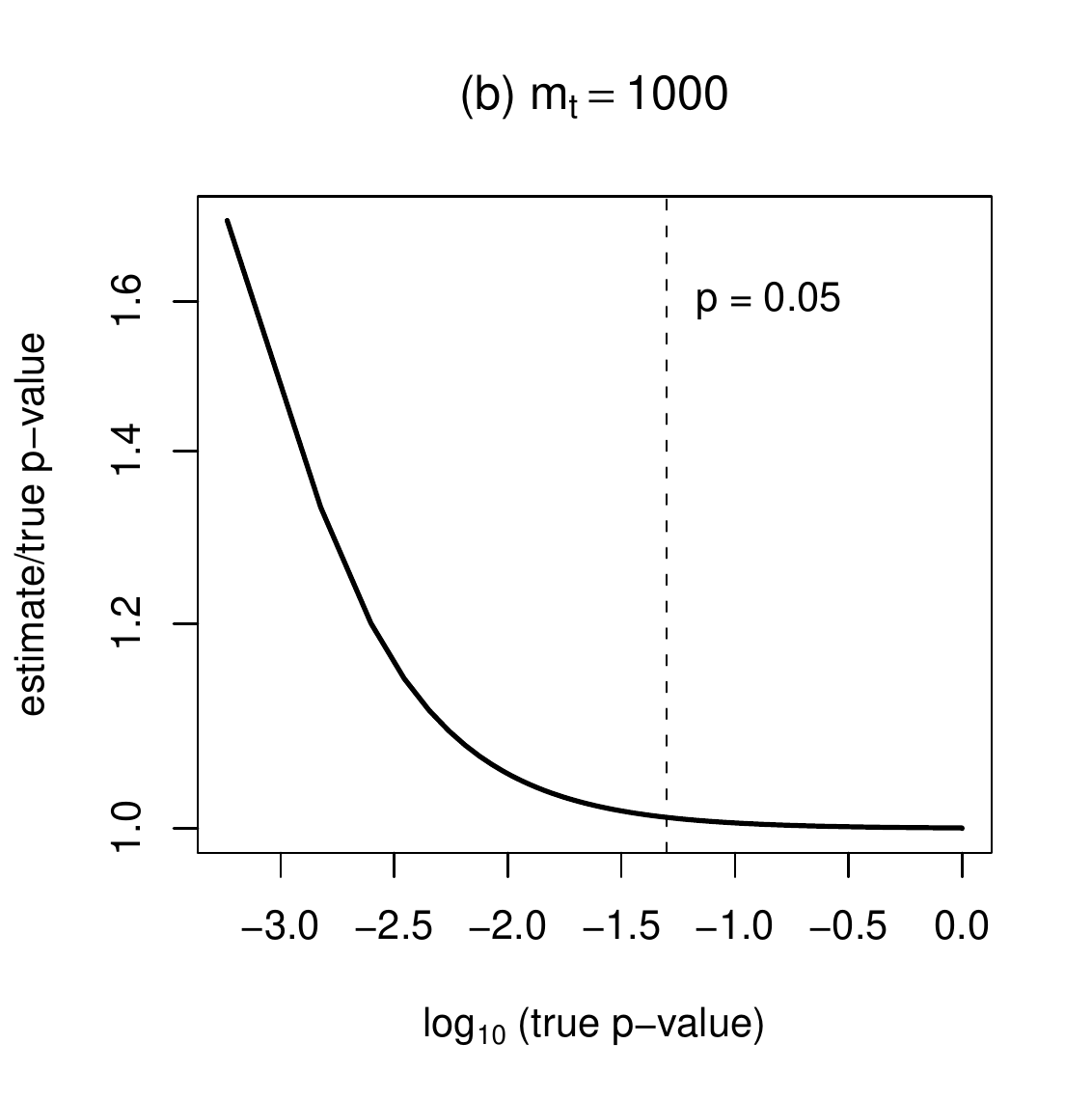}
\caption{
Ratio of the upper bound $p_u=(b+1)/(m+1)$ to the exact $p$-value.
Panel (a) is for $m_t=126$ and (b) is for $m_t=1000$.
The number of permutations is $m=1000$ in each case, and values are plotted for all possible values of $b$.
}
\label{fig:puvspe}
\end{figure}

\subsection{Specific scenarios}

The number $m_t$ of possible distinct values of the test statistic is not simply a function of the number of permutations, but also depends on the form of the test statistic and on the sample sizes.
This is because different permutations can lead to the same test statistic, especially for two-sided tests with balanced samples.

Consider a comparison between two groups of sizes $n_1$ and $n_2$.
In this situation it is usual to sample permutations by randomly assigning $n_1$ of the indices $i$ to group 1 and the remaining $n_2$ to group 2.
If the test statistic is one-sided, or if $n_1\ne n_2$, then each permutation will generally yield a distinct test statistic.
In that case, the number of distinct values of the permuted statistics is
$$m_t+1={n_1+n_2\choose n_1}.$$
For example, a one-sided test with $n_1=n_2=5$ gives $m_t+1={10 \choose 5}=252$.
The integral approximation (\ref{eq:two}) provides an excellent approximation to the exact $p$-values (Table~\ref{tab:twogroups}).

\begin{table}
\caption{Exact and approximate $p$-values for a one-sided test between two groups of size five using $m=100$ random permutations.}
\vspace{6pt}
\centering
\begin{tabular}{ccccc}
\hline
Number of & Estimated & Exact & Integral approximated & Upper bound\\
$t_{\rm sim}\ge t_{\rm obs}$ & $\hat p$-value & $p$-value & $p$-value & $p_u$\\
\hline
0 & 0 & 0.008047755 & 0.008101416 & 0.0099 \\
1 & 0.01 & 0.017818517 & 0.017829558 & 0.0198 \\
2 & 0.02 & 0.027718516 & 0.027719402 & 0.0297 \\
3 & 0.03 & 0.037619829 & 0.037619855 & 0.0396 \\
4 & 0.04 & 0.047520825 & 0.047520824 & 0.0495 \\
5 & 0.05 & 0.057421814 & 0.057421814 & 0.0594 \\
6 & 0.06 & 0.067322804 & 0.067322804 & 0.0693 \\
7 & 0.07 & 0.077223794 & 0.077223794 & 0.0792 \\
\vdots & \vdots & \vdots & \vdots & \vdots \\
\hline
\end{tabular}
\label{tab:twogroups}
\end{table}

If a two-sided test statistic is used, for example the absolute value of a $t$-statistic, and if the two groups are of equal size $n_1=n_2$, then each permutation gives the same value of the test statistic as the corresponding permutation with the two groups reversed.
In this case, the total number of distinct test statistics possible is $m_t+1={n_1+n_2\choose n_1}/2$.

If there are three or more groups of unequal sizes then $m_t+1$ is the multinomial coefficient
\[
m_t+1={\sum_{j=1}^k n_j\choose n_1\,\ldots\,n_k}
\]
where $k$ is the number of groups.
If some of the group sizes $n_j$ are equal, then some permutations will in general give equal test statistics, so the calculaton of $m_t$ will need to consider the particular configuration of samples sizes.

\section{Discussion and recommendations}

The approach posited in Section~5, of drawing permutations without repeating previous values of the test statistic, turns out to be a surprisingly difficult combinatorial programming exercise.
Hence this approach is seldom used---so much so that we have never seen it in the genomic literature.
Nevertheless, this is the superior approach to permutation tests.
Sampling without replacment gives rise to a simple easily understood exact $p$-value in the form of $p_u$.
Even more importantly, it is possible to show that sampling permutations without replacement gives more statistical power than sampling with replacement for any given number of permutations $m\le m_t$.
Our first recommendation then is that more attention be given to this problem by computer scientists and combinatorialists, to provide biostatisticians with efficient numerical algorithms to sample permutations without replacement.

Our second recommendation is that ``unbiased'' estimators of $p$-values such as $\hat p$ should not be used, because they can't be guaranteed to control the type I error rate.
Although the risk involved in using $\hat p$ in place of the exact $p$-value may be modest in many applications, there is no justification for using an invalid $p$-value when a valid $p$-value, in the form of $p_u$, is readily available for no extra cost.

Our third recommendation is that users might consider using the exact $p$-value $p_e$, at the cost of a trivial amount of extra computation.
Our algorithm to compute $p_e$ is implemented as the function permp in the software package statmod for R available from \url{http://www.r-project.org}.
The exact value $p_e$ gives a worthwhile gain in statistical power over $p_u$ when the number of permutations $m$ performed is a more than negligible proportion of the total number possible $m_t$.

A referee asked us to warn users that the number of permutations $m$ should not be chosen too small.
While this is true, the urgent need to avoid small $m$ disappears when the exact $p$-value $p_e$
is used in place of $\hat p$, because $p_e$ ensures a valid statistical test regardless of sample size or the number of permutations.
When exact $p$-values are used, the only penalty for choosing $m$ small is a loss of statistical power to reject the null hypothesis.
This is as it should be: more permutations should generally provide greater power.

The same referee asked us to comment on the difference between permuting genes and permuting samples.
One obvious difference between genes and samples is that there are generally more of the former than the latter, so that $m_t$ is likely to be astronomically large when genes are permuted.
This implies that $p_u$ is likely to be a good approximation for $p_e$ in this context.
We have avoided discussing gene permutation in this article however because gene permutation has another obvious problem.
Permutation assumes that the items being permuted are statistically independent, whereas genes usually cannot be assumed independent in genomic applications (Goeman and B\"uhlman, 2007; Wu et al, 2010).
In general, we recommend permuting samples rather than permuting genes when there is a choice.

Finally, it is worthwhile mentioning that there are alternatives to permutation that alleviate the problem of small $m_t$ when the number of samples is not large (D{\o}rum, 2009; Wu et al, 2010).

\begin{thebib}

\item Barnard, G.A. (1963) Discussion of ``The spectral analysis of point processes" by MS Bartlett. \textit{J.R. Statist. Soc.} B, \textbf{25}, 294.

\item Davison, A.C. and Hinkley, D.V. (1997) \textit{Bootstrap methods and their application}, Cambridge University Press, United Kingdom.

\item Dwass, M. (1957) Modified randomization tests for nonparametric hypotheses. \textit{Ann. Math. Statist.}, \textbf{28}, 181--187.

\item D{\o}rum, G., Snipen, L., Solheim, M., and Saebo, S. (2009). Rotation testing in gene set
enrichment analysis for small direct comparison experiments. \textit{Stat. Appl. Genet. Mol.
Biol}, \textbf{8}, Article 34.

\item Efron, B. (1982) \textit{The jackknife, the bootstrap and other resampling plans}, Society for Industrial and Applied Mathematics, England.

\item Efron, B. and Tibshirani, R.J.(1993) \textit{An introduction to the bootstrap}, Chapman \& Hall, New York.

\item Edgington, E.S. and Onghena, P. (2007) \textit{Randomization tests}, Fourth Edition, Chapman \& Hall/CRC, Boca Raton, FL. 

\item Ernst, M.D. (2004) Permutation methods: a basis for exact inference. \textit{Statistical Science}, \textbf{19}, 676--685.

\item Fisher, R.A. (1935) \textit{The Design of Experiments}, 3rd Edition, Oliver \& Boyd, London.

\item Goeman, J.J. and B\"uhlmann, P. (2007) Analyzing gene expression data in terms of gene sets: methodological issues. \textit{Bioinformatics}, \textbf{23}, 980--987.

\item Good, P. (1994) \textit{Permutation tests: a practical guide to resampling methods for testing hypotheses}, Springer-Verlag New York, New York.


\item Higgins, J.J. (2004) \textit{An introduction to modern nonparametric statistics}, Brooks/Cole, Pacific Grove, CA.

\item Holm, S. (1979). A simple sequentially rejective multiple test procedure. \textit{Scandinavian Journal of Statistics}, \textbf{6}, 65--70.


\item Manly, B.F.J. (1997) \textit{Randomization, bootstrap and Monte Carlo methods in biology}, Second Edition, Chapman \& Hall, London.

\item Mielke, P.W. and Berry, K.J. (2001) \textit{Permutation tests: a distance function approach}, Springer-Verlag New York, New York. 

\item Onghena, P. and May, R.B. (1995) Pitfalls in computing and interpreting randomization test $p$ values: A commentary on Chen and Dunlap. \textit{Behaviour Research Methods, Instruments, \& Computers}, \textbf{27}, 408--411.

\item Wu, D., Lim, E., Vaillant, F., Asselin-Labat, M., Visvader, J.E. and Smyth, G.K. (2010) ROAST: rotation gene set tests for complex microarray experiments. \textit{Bioinformatics}, \textbf{26}, 2176--2182. 

\end{thebib}

\end{document}